\documentclass[floatfix,aps,twocolumn,superscriptaddress,pra,10pt]{revtex4-1}

\usepackage[T1]{fontenc}
\usepackage{amssymb}
\usepackage{amsmath}
\usepackage{multirow}
\usepackage[dvips]{graphicx}
\usepackage{bm}
\usepackage{float}
\usepackage{color}
        \usepackage{tabularx}

\graphicspath{{figs/}}

\newcommand{\ul}[1]{\bm{#1}}

\newcommand{\idkk}{\bm{1}_{\mathcal{K}(\vec{k})}}
\newcommand{\idc}{\bm{1}_{\mathcal{C}}}
\newcommand{\beq}{\begin{equation}}
\newcommand{\eeq}{\end{equation}}
\newcommand{\bea}{\begin{eqnarray}}
\newcommand{\eea}{\end{eqnarray}}

\newcommand{\al}{\alpha}

\newcommand{\iom}{i\omega_n}

\newcommand{\be}{\beta}

\newcommand{\ga}{\gamma}

\newcommand{\up}{\uparrow}
\newcommand{\dn}{\downarrow}

\newcommand{\ket}[1]{| #1 \rangle}
\newcommand{\bra}[1]{\langle #1 |}

\begin{document}
\title{Spin-orbit and anisotropic strain effects on the electronic correlations of Sr$_2$RuO$_4$}

\author{Jorge I. Facio}
\affiliation{Centro At{\'o}mico Bariloche and Instituto Balseiro, CNEA, CONICET, (8400) Bariloche, Argentina}
\affiliation{Institute for Theoretical Solid State Physics, IFW Dresden, Helmholtzstrasse 20, 01069 Dresden, Germany}
\author{Jernej Mravlje}
\affiliation{Jo\v{z}ef Stefan Institute, Jamova 39, SI-1000 Ljubljana, Slovenia}
\author{Leonid Pourovskii}
\affiliation{CPHT, Ecole Polytechnique, CNRS,  Universit\'e Paris-Saclay, Route de Saclay, 91128 Palaiseau, France}
\affiliation{Coll\`{e}ge de France, 11 place Marcelin Berthelot, 75005 Paris, France}
\author{Pablo S. Cornaglia}
\affiliation{Centro At{\'o}mico Bariloche and Instituto Balseiro, CNEA, CONICET, (8400) Bariloche, Argentina}
\author{V. Vildosola}
\affiliation{Departamento de Materia Condensada, GIyA, CNEA, CONICET, (1650) San Mart\'{\i}n, Provincia de Buenos Aires, Argentina}

\begin{abstract}
We present an implementation of the rotationally invariant slave boson technique as an impurity solver for density functional 
theory plus dynamical mean field theory (DFT+DMFT). Our approach provides explicit relations between quantities in the local correlated subspace treated with DMFT and the Bloch basis used to solve the DFT equations. In particular, we present an expression for the mass enhancement of the quasiparticle states in reciprocal space.
We apply the method to the study of the electronic correlations in Sr$_2$RuO$_4$ under anisotropic strain. 
We find that the spin-orbit coupling plays a crucial role in the mass enhancement differentiation between the quasi-one-dimensional $\al$ and $\be$ bands, and on its momentum dependence over the Fermi surface.
The mass enhancement, however, is only weakly affected by either uniaxial or biaxial strain, even accross the Lifshitz transition induced by the strain.
\end{abstract}

\maketitle

\section{Introduction}
\label{sec:intro}
The electronic properties of transition metal oxides continue to be a central problem in condensed matter physics.
Part of the challenge is due to the sensitivity of the low energy physics of these systems to the complex interplay between 
the crystal structure, the amount of hybridization between oxygens and the transition metal \cite{cava2004schizophrenic},
the Coulomb interaction, including the correct differentiation between intraorbital e interorbital interactions \cite{mravlje2011coherence,PhysRevB.93.155103}, and the spin-orbit coupling (SOC) \cite{PhysRevLett.85.5194,zhang2016fermi}.
This interplay seems to be the key to understanding the normal state, magnetic, and superconducting properties of some of these compounds \cite{Tokura462}. In particular, the role of the SOC and of the proximity to a Lifshitz transition have recently attracted much interest.

It has been reported that the SOC is crucial to explaining the insulating character of Sr$_2$IrO$_4$ \cite{PhysRevLett.107.266404}, and the magnetic properties of Ca$_2$RuO$_4$ \cite{PhysRevB.95.075145} and Sr$_3$Ru$_2$O$_7$ \cite{behrmann2012multiorbital}. It may also be relevant to determine the nature of the superconducting state in Sr$_2$RuO$_4$ \cite{veenstra2014spin,ayarcha}. At the local density approximation (LDA) level, its inclusion is necessary to improve the Fermi surface shape of Sr$_2$RuO$_4$ and Sr$_2$RhO$_4$ as compared to ARPES data~\cite{PhysRevB.74.035115,zhang2016fermi,kim2017spin,PhysRevLett.85.5194,PhysRevLett.101.026406}.


Recently, the possibility of inducing a Lifshitz transition in Sr$_2$RuO$_4$ by applying external strains has been addressed experimentally \cite{steppke2016strong,burganov2016strain}.
Single crystals under a uniaxial strain applied in the [100] direction present a peak in the superconducting critical temperature as a function of the stress. 
The peak position seems to coincide with the value of the stress at which the Lifshitz transition takes place and the associated Van Hove singularity (VHS) crosses the Fermi level \cite{hicks2014strong,steppke2016strong}. 
This transition has been observed in ARPES measurements of epitaxial thin films of Sr$_2$RuO$_4$ and Ba$_2$RuO$_4$ grown over different substrates  \cite{burganov2016strain}.
Different values of the lattice parameter $a$ are obtained adjusting the lattice mismatch and the results can be interpreted as the behavior of Sr$_2$RuO$_4$ under a biaxial strain in the [100] and [010] directions.

These experiments have triggered theoretical investigations focused on understanding how the Lifshitz transition
affects the pairing properties in the superconducting phase \cite{hsu2016manipulating,hsu2017band} or the spin susceptibility in the normal phase \cite{PhysRevB.94.224507}. 
An interesting open question is to what extent this transition affects the electronic correlations in the normal phase.
Indeed, the proximity of the VHS to the Fermi level has been found to be important to understand the anisotropic mass 
enhancement of quasiparticles. The Fermi surface of this material is composed by the sheets $\al$, $\be$ and $\ga$ of mainly Ru $t_{2g}$ character, whose respective renormalized masses, $m^{*}/m_{LDA}$, have been reported to be 3, 3.5, and 5, respectively~\cite{mackenzie2003superconductivity}. 
Density functional theory suplemented by dynamical mean field theory (DFT+DMFT) calculations, have shown that the larger renormalization of the $\ga$ sheet can be associated with the proximity of the VHS to the Fermi level, and Hund's rule coupling effects \cite{mravlje2011coherence}.

The purpose of this article is twofold. First, to introduce an implementation of the rotationally invariant slave-boson (RISB) method \cite{li1989spin,Lechermann2007,Ferrero2009,Ferrero2009a,piefke2017rigorous} as an impurity solver for DFT+DMFT. The RISB method is a low weight numerical method geared to describe Fermi liquid behavior and has been successfully used, supplemented by DFT calculations, to describe the low energy correlations of materials \cite{lechermann2009correlation,behrmann2012multiorbital,grieger2010electronic,behrmann2015interface,piefke2011lda+,schuwalow2010realistic}. 
Our approach provides an explicit relation between the mass enhancement calculated in the quantum impurity problem and the mass enhancement that different quasiparticle states acquire after embedding the impurity self-energy back in the lattice problem.
This relation can also be applied when using other techniques to solve the quantum impurity problem.
Second, we apply this methodology to Sr$_2$RuO$_4$. For the unstressed compound, using the above mentioned relation, we explain why 
Bloch states on the $\al$ and $\be$ sheets of the Fermi surface, having mostly $xz$ and $yz$ character (which correspond in the quantum impurity problem to degenerated cubic Wannier orbitals), acquire a different renormalization, as experimentally observed. 
To finalize, motivated by the experiments of Ref. \cite{steppke2016strong,burganov2016strain}, we analyze how the electronic correlations, as measured by the mass enhancement, evolve under biaxial or uniaxial stress, and to what extent they are affected by the Lifshitz transition.

The rest of this paper is organized as follows.
In Section \ref{sec:methods} we introduce our implementation of RISB as an impurity solver for DFT+DMFT.
In Section \ref{sec:sr2ruo4} we analyze how the method describes the correlated metal Sr$_2$RuO$_4$. 
In Section \ref{sec:anisotropic} we present results for the material under anisotropic stress.
Finally, in Section \ref{sec:conclusion} we present our concluding remarks.

\section{Method: DFT+DMFT using RISB as impurity solver}
\label{sec:methods}

In this section we first outline the DFT+DMFT scheme as implemented in Refs. \cite{aichhorn2009dynamical,aichhorn2016triqs} in order to define the notation. We next describe the implementation of the RISB method as a multiorbital impurity solver and its use for DMFT calculations.  Finally, we derive a relation between quantities in the correlated subspace and the Bloch space which allows to determine the mass renormalization of the Bloch states.

\subsection{DFT+DMFT scheme.}

The first step is the solution of the DFT Kohn-Sham equations, which yield  
the Kohn-Sham energies, $\varepsilon_{\vec{k}\nu}$, and the corresponding states, $|\psi_{\vec{k},\nu}\rangle$, 
classified by the crystal momentum $\vec{k}$ and a band index $\nu$.
The second step is the treatment of the strong local correlations using DMFT.
To that aim, a set of Wannier orbitals $|\chi^{\vec{R}}_{m}\rangle$ is constructed, where $\vec{R}$ labels a lattice site and $m$ denotes the orbital and spin degrees of freedom 
\footnote{In this work we use the projective technique implemented in Ref. \cite{PhysRevB.77.205112}.}.
We define $\mathcal{C}$ as the space spanned by the set of correlated Wannier orbitals at a given site, and we omit the site label in the following. 
In the so-called projective method, only Bloch bands whose energy lies within a predefined energy window $\mathcal{W}$ are used in the construction of the Wannier orbitals.
We define $\mathcal{K}$ as the space spanned by all the Bloch states whose energy lies within $\mathcal{W}$, and $\mathcal{K}(\vec{k})$ as
a subspace of $\mathcal{K}$ formed by states with a definite crystal momentum $\vec{k}$.
We also define $\ul{P}(\vec{k})$ as the transformation operator from $\mathcal{K}(\vec{k})$ to $\mathcal{C}$
whose matrix elements are $P_{m,\nu}(\vec{k}) = \langle \chi_{m} | \psi_{\vec{k},\nu}\rangle$.
Denoting by $\idc$ and $\idkk$ the identity matrices in $\mathcal{C}$ and $\mathcal{K}(\vec{k})$ respectively, these transformations satisfy: $\ul{P}(\vec{k}) \ul{P}^\dag(\vec{k}) = \idc$,
but the converse ($\ul{P}^\dag(\vec{k}) \ul{P}(\vec{k}) = \idkk$) only is fulfilled if the number of bands at $\vec{k}$ and within $\mathcal{W}$ is equal to the number of Wannier orbitals in $\mathcal{C}$.

The lattice Green's function in Matsubara representation reads:
\begin{equation}
\label{eq:gkw}
\ul{G}^{-1}(\vec{k},\iom) = \big( \iom +\mu \big) \idkk - \ul{\varepsilon}(\vec{k}) - \ul{\Sigma}(\vec{k},\iom),
\end{equation}
where $\ul{\varepsilon}(\vec{k})$ is a diagonal matrix formed by the Kohn-Sham eigenvalues of all the bands within $\mathcal{W}$ at a given $\vec{k}$ point, and $\ul{\Sigma}(\vec{k},\iom)$ is constructed by embedding in $\mathcal{K}$ a local self-energy, $\ul{\Sigma^{imp}}(\iom)$, 
calculated through an auxiliary quantum impurity problem introduced by DMFT:
\begin{equation}
\label{eq:sigmak}
\ul{\Sigma}(\vec{k},\iom) = \ul{P}^\dag(\vec{k}) \Big( \ul{\Sigma^{imp}}(\iom) - \ul{\Sigma^{dc}} \Big) \ul{P}(\vec{k}),
\end{equation}
where $\ul{\Sigma^{dc}}$ is a correction included to reduce the double counting of interactions in the DFT+DMFT method.

The quantum impurity problem consists of a local term, $H^{loc}$, which includes one-body energies and interaction terms in $\mathcal{C}$; and an hybridization term, which describes the coupling of these Wannier orbitals to an effective non-interacting fermionic bath, which is determined selfconsistently.
The local Hamiltonian reads
\begin{equation}
H^{loc} = \sum_{mm'}\varepsilon^0_{mm'}d^\dag_m d_{m'} + H^{int},
\end{equation}
where $H^{int}$ describes the interactions, and $\ul{\varepsilon^0}$ are the one-body energies, which we compute as
\begin{equation}
    \label{eq:obe}
    \ul{\varepsilon^0} = \frac{1}{\mathcal{N}}\sum_{\vec{k}}\ul{P}(\vec{k}) \ul{\varepsilon}(\vec{k}) \ul{P}^\dag(\vec{k}) - \ul{\Sigma^{dc}}.
\end{equation}
where $\mathcal{N}$ is the number of $k$-points.
The effective bath is described by an hybridization function, $\ul{\Delta}(\iom)$, which is determined at each step of the DMFT cycle. 

From the lattice Green's function [see Eq. (\ref{eq:gkw})] we define a local Green's function in $\mathcal{C}$
\begin{equation}
\label{eq:Glocproj}
\ul{G^{loc}}(\iom) = \sum_{\vec{k}} \ul{P}(\vec{k}) \ul{G}(\vec{k},\iom) \ul{P}^\dag(\vec{k}),
\end{equation}
while the impurity Green's function of the DMFT auxiliary problem reads
\begin{equation}
\label{eq:Gqip}
\ul{G^{imp}}(\iom) = [(\iom+\mu)\idc - \ul{\varepsilon^0} - \ul{\Delta}(\iom) - \ul{\Sigma^{imp}}(\iom)]^{-1},
\end{equation}
where $\ul{\Sigma^{imp}}(\iom)$ is determined solving the auxiliary quantum impurity problem for a given $\ul{\Delta}(\iom)$.  
The DMFT self-consistency is fulfilled for a $\ul{\Delta}(\iom)$ such that $\ul{G^{imp}}= \ul{G^{loc}}$.


\subsection{RISB as impurity solver}

In the RISB formalism \cite{li1989spin,Lechermann2007,Ferrero2009,Ferrero2009a,facio2017nature} 
the physical fermionic operators $d_{m}$, which destroy an electron in the Wannier state $|\chi_{m}\rangle$, are represented as a linear combination of an equal number of auxiliary fermionic operators $f_{m}$:
\begin{equation}
\label{eq:rel_df}
d_{m} = \sum_{m^\prime}R_{\mathcal{C}mm'}[\{\phi_{AB}\}] f_{m'}.
\end{equation}
Here, the matrix $\mathbf{R}_{\mathcal{C}}$ is a function of a set of auxiliary boson fields $\{\phi_{AB}\}$, where the indices $A$ and $B$ refer to the local multiplets, and describes
the different processes by which an electron can be destroyed.
Its form ensures that the matrix elements of the $d_{m}$ operators remain the same in the new representation \cite{Lechermann2007}.

In the enlarged Hilbert space, spanned by the auxiliary fermion and boson fields, $H^{loc}$ is represented as a quadratic
form in the auxiliary boson operators:
\begin{equation}
\label{eq:rep_hat}
H^{loc} = \sum_{AB} \langle A | H^{loc} | B \rangle \sum_{n}\phi_{An}^\dag \phi_{Bn}. 
\end{equation}
A one-to-one mapping with the original local Hilbert space is obtained with the introduction of time-independent Lagrange multipliers $\lambda_0$ and ${\Lambda}_{\mathcal{C}mm^\prime}$ that enforce the following constraints:
\begin{eqnarray} \label{eq:constr0}
\sum_{A,B,C}\phi_{CA}^\dagger\phi_{CB}\langle B |\hat{O}|A\rangle&=&\hat{O},
\end{eqnarray}
where $\hat{O}=\{1,\, f^\dagger_m f_{m'}\}$.

In the saddle-point approximation, the boson fields are replaced by classical numbers, and the self-energy acquires a simple form \begin{equation}
\label{eq:sigmaRISB}
\ul{\Sigma^{imp}}(\iom) = \iom \ul{\Sigma^{imp}}_{1} + \ul{\Sigma^{imp}}_{0},
\end{equation}
where
\begin{align}
\label{eq:s1}
\ul{\Sigma^{imp}}_{1} &= \ul{1}_\mathcal{C} - [\ul{R}_\mathcal{C}\ul{R^\dag}_\mathcal{C}]^{-1}, 
\end{align}
effectively renormalizes the hybridization with the non-interacting bath, and
\begin{align}
\label{eq:s0}
\ul{\Sigma^{imp}}_{0} &= \Big(\ul{1}_\mathcal{C} - [\ul{R}_\mathcal{C}\ul{R^\dag}_\mathcal{C}]^{-1} \Big) \mu +  \ul{R^{\dag-1}}_\mathcal{C} \mathbf{\Lambda}_{\mathcal{C}} \ul{R^{-1}}_\mathcal{C}-\ul{\varepsilon_0},
\end{align}
renormalizes the level positions.

Replacing the self-energy of Eq. (\ref{eq:sigmaRISB}) in Eq. (\ref{eq:Gqip}) the impurity Green's function reads
\begin{equation}
    \ul{G^{imp}} = \ul{R}_\mathcal{C}\Big[\big(\iom+\mu\big)\ul{1}_\mathcal{C} - \mathbf{\Lambda}_\mathcal{C} - \ul{R^\dag}_\mathcal{C} \ul{\Delta}(\iom) \ul{R}_\mathcal{C}\Big]^{-1} \ul{R^\dag}_\mathcal{C}.
\end{equation}
The auxiliary fermionic fields $f_m$ can be interpreted as quasiparticle degrees of freedom with a quasiparticle weight $\ul{Z}_\mathcal{C} = \ul{R}_\mathcal{C} \ul{R^\dag}_\mathcal{C}$. Their associated Green's function is [using Eq. (\ref{eq:rel_df})]
\begin{equation}
    \ul{G_{qp}^{imp}} = \Big[\big(\iom+\mu\big)\ul{1}_\mathcal{C} - \mathbf{\Lambda}_\mathcal{C} - \ul{R^\dag}_\mathcal{C} \ul{\Delta}(\iom) \ul{R}_\mathcal{C}\Big]^{-1}. 
\end{equation}

\subsection{RISB method in the DFT+DMFT scheme}
When the RISB technique is used to solve DMFT's impurity problem, the relations between physical and quasiparticle quantities (e.g. operators and correlators) introduced in the subspace $\mathcal{C}$ are expected to have analogues in $\mathcal{K}$. 
In particular, the physical fermionic operator $c^\dag_{\vec{k}\nu}$ which creates an electron in the Kohn-Sham state $|\psi_{\vec{k},\nu}\rangle$ can be related to quasiparticle operators $\tilde{c}_{k\nu}$ through transformation matrices $R_{\mathcal{K}\nu\nu'}(\vec{k})$:
\begin{equation}
\label{eq:rel_dfk}
c_{\vec{k}\nu} =\sum_{\nu^\prime} R_{\mathcal{K}\nu\nu'}(\vec{k}) \tilde{c}_{\vec{k}\nu'}.
\end{equation}
and accordingly, the lattice Green's function [see Eq. (\ref{eq:gkw})] be written in terms of the quasiparticle lattice Green's function, $\ul{G_{qp}}(\vec{k},\iom)$:
\begin{equation}
\label{eq:relation}
\ul{G}(\vec{k},\iom) = \ul{R}_\mathcal{K}(\vec{k}) \ul{G_{qp}}(\vec{k},\iom) \ul{R}_\mathcal{K}^\dag(\vec{k}).
\end{equation}


In the RISB saddle-point approximation the lattice self-energy reads: 
\begin{align}
\label{eq:lio}
\ul{\Sigma}(\vec{k},\iom) &= \iom \ul{P}^\dag(\vec{k}) \ul{\Sigma^{imp}}_{1} \ul{P}(\vec{k}) \nonumber \\&+ \ul{P}^\dag(\vec{k}) \big[ \ul{\Sigma^{imp}}_{0}-\ul{\Sigma}^{dc} \big] \ul{P}(\vec{k}),
\end{align}
which leads to
\begin{equation*}
    \ul{P}^\dag(\vec{k}) \ul{\Sigma^{imp}}_{1} \ul{P}(\vec{k}) =  \idkk - \ul{Z}_\mathcal{K}^{-1}(\vec{k}),
\end{equation*}
and using Eq. (\ref{eq:s1}) to
\begin{equation}
\label{Znu}
\ul{Z}^{-1}_\mathcal{K}(\vec{k}) = \idkk-\ul{P}^\dag(\vec{k})\ul{P}(\vec{k}) + \ul{P}^\dag(\vec{k}) \ul{Z}_{\mathcal{C}}^{-1}\ul{P}(\vec{k}),
\end{equation}
that accounts for the mass renormalization of the Bloch states due to the electronic correlations \footnote{The inverse relations are presented in Appendix \ref{app:inverse}.}. 

Equation (\ref{Znu}) gives the quasiparticle weight in momentum space in terms of the quasiparticle weights obtained from DMFT's auxiliary quantum
impurity problem.
Note that this relation was obtained using general assumptions for the low energy behavior of the self-energy in the Bloch and Wannier basis and can therefore be used independently of the quantum impurity solver employed to calculate $\ul{Z}_{\mathcal{C}}$.
Within this theory, the $\vec{k}$ dependence of the mass enhancement stems from the different amplitudes the Bloch states have in the Wannier states of the correlated subspace $\mathcal{C}$.

The quasiparticle lattice Green's function can be calculated from Eqs. (\ref{eq:gkw}) and (\ref{eq:relation}) (see Appendix \ref{ap_gqp}). We have:
\begin{equation}
    \ul{G^{-1}_{qp}}(\vec{k},\iom) = \big(\iom+\mu\big) \ul{1}_\mathcal{K}-\ul{\Lambda}_{\mathcal{K}}(\vec{k})-\ul{R}_\mathcal{K}^\dag(\vec{k}) \ul{\tilde{\varepsilon}}(\vec{k})  \ul{R}_\mathcal{K}(\vec{k}),
\label{eq:gqp} 
\end{equation}
where $\ul{\Lambda}_\mathcal{K}(\vec{k})= \ul{P}^\dag(\vec{k}) \ul{\Lambda}_\mathcal{C} \ul{P}(\vec{k})$ and  
 $ \ul{\tilde{\varepsilon}}(\vec{k}) = \ul{\varepsilon}(\vec{k})-\ul{P}^\dag(\vec{k})\big(\ul{\varepsilon^0}+\ul{\Sigma^{dc}}\big)\ul{P}(\vec{k})$.
 From the definition of $\ul{\varepsilon^0}$ [see Eq. (\ref{eq:obe})] it follows that the energies $\ul{\tilde{\varepsilon}}(\vec{k})$ satisfy $\sum_{\vec{k}}\ul{P}(\vec{k}) \ul{\tilde{\varepsilon}}(\vec{k}) \ul{P}^\dag(\vec{k}) = 0$, and the level energies are therefore controlled by $\ul{\Lambda}_\mathcal{K}(\vec{k})$ while the transformation matrix 
 $\ul{R}_\mathcal{K}(\vec{k})$ renormalizes the bandwidth. In Appendix \ref{app:inverse} we show that $\ul{R}_\mathcal{C}$ and $\ul{R}_\mathcal{K}$ can be related by an equation analogous to Eq. (\ref{Znu}).

The quasiparticle wave-functions $\ket{\psi^{QP}_{\vec{k},\nu}}$ can be obtained as the eigenvectors of $\ul{{G}_{qp}}^{-1}(\vec{k},\iom\to0)$, 
and are linear combinations of the Kohn-Sham basis provided by the DFT calculation, $\ket{\psi_{\vec{k},\nu}}$.
\begin{equation}
\ket{\psi^{QP}_{\vec{k},\nu}} = \sum_{\nu^\prime}  \mathcal{U}^\dag_{\nu,\nu^\prime}(\vec{k}) \ket{\psi_{\vec{k},\nu^\prime}}.
\end{equation}
The mass renormalization of the quasiparticle states is obtained applying the unitary transformation $\mathcal{U}$ to the low energy expansion of the lattice self-energy function [given by Eq. (\ref{eq:sigmak})], which leads to
\begin{equation}
    \ul{Z}_\mathcal{K}^{QP}(\vec{k}) = \mathcal{U}(\vec{k}) \ul{Z}_\mathcal{K}(\vec{k})\mathcal{U}^\dag(\vec{k}).   
    \label{eq:ZQP}
\end{equation}

A point $\vec{k}_F$ belongs to the renormalized Fermi surface $\mathcal{S}$ if there is a state $\ket{\psi^{QP}_{\vec{k}_F,\nu_F}}$ which is a zero energy eigenvector  of the function $\ul{{G}_{qp}}^{-1}(\vec{k}_F,\iom\to0)$.
In the general case, the matrix $\ul{Z}_\mathcal{K}^{QP}(\vec{k})$ is non-diagonal. We may, however estimate the quasiparticle weight associated with a Fermi sheet $\nu_F$ by projecting $\ul{Z}_\mathcal{K}(\vec{k})$ onto the corresponding quasiparticle states at the Fermi level:
\begin{equation}
    Z_{\nu_F}^{QP}(\vec{k}_F)=\bra{\psi^{QP}_{\vec{k}_F,\nu_F}}\ul{Z}_\mathcal{K}(\vec{k}_F)\ket{\psi^{QP}_{\vec{k}_F,\nu_F}}.
    \label{eq:ZQPFS}
\end{equation}
The quasiparticle weight $\ul{Z}^{QP}_\mathcal{K}(\vec{k})$ provides information about the mass renormalization obtained at particular line cuts of the Brilloin zone, which can be obtained performing Angle Resolved Photoemission (ARPES) or dHvA oscillations experiments.

As a benchmark of the method, in Appendix \ref{svo_results} we present a comparison of results obtained by solving the quantum impurity problem with RISB or with CTQMC for the cubic perovskite SrVO$_3$.
\section{Application to S\MakeLowercase{r$_2$}R\MakeLowercase{u}O$_4$}
\label{sec:sr2ruo4}
The ruthenate Sr$_2$RuO$_4$ was analyzed in Ref. \cite{mravlje2011coherence} using LDA+DMFT(CTQMC) in the absence of SOC and more recently in Ref.  \cite{kim2017spin} including the SOC at the DFT level. These studies show that the Hund's rule coupling plays a central role in explaining the magnitude of the quasiparticle weight and its orbital differentiation. The magnitude of the latter being also affected by the presence of a Van Hove's singularity near the Fermi level. These results present a remarkable agreement with the experimental effective masses and with ARPES and NMR data. 
Although including the SOC reduces the degeneracy of the local multiplet structure it does not change the Hund's correlated metal nature of the compound.

\begin{figure}[ht]
\center
\includegraphics[width=0.4\textwidth,angle=0,keepaspectratio=true]{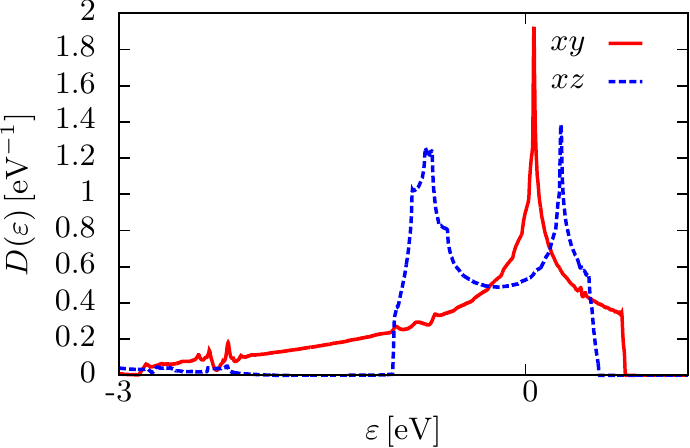}
\caption{LDA density of states of Sr$_2$RuO$_4$ projected on the Ru-$t_{2g}$ orbitals and within the energy window $\mathcal{W}_s$ considered in this work to construct the Wannier orbitals.}
\label{sro_ws}
\end{figure}

We apply below the DFT+DMFT(RISB) method detailed in the previous section to analyze the electronic structure of Sr$_2$RuO$_4$.  
We take the experimental crystal structure extracted from Ref. \cite{PhysRevB.57.5067} with lattice parameters $a=3.862\,\text{\AA}$, $c=12.722\,\text{\AA}$.
We use the \textsc{wien2k} code \cite{wien2k} with the \textsc{triqs} interface for DFT+DMFT \cite{aichhorn2009dynamical,aichhorn2016triqs,triqs_wien2k_full_charge_SC} and consider the Local Density Approximation (LDA) for the exchange and correlation potential at the DFT level using a dense $k$ mesh of $39\times39\times39$ points \footnote{We use this dense mesh in order to address the changes in the electronic structure associated to the Van Hove singularity in particular for Section \ref{sec:anisotropic}.}.
To construct the Wannier orbitals we take an energy window, $\mathcal{W}_s = [-3\,\text{eV},1.3\,\text{eV}]$, which basically contains the $t_{2g}$ bands of the Ru atom as indicated in Fig. \ref{sro_ws}.

The Coulomb interaction within the $t_{2g}$ manifold is described with the rotationally invariant Kanamori Hamiltonian:
\begin{eqnarray} \label{eq:kanamori}
        H^{int} &=& U \sum_{m} n_{m\up} n_{m\dn} + (U-2J) \sum_{m\neq m^\prime} n_{m\up} n_{m^\prime\dn} + \nonumber\\
&+& (U-3J) \sum_{m>m^\prime,\sigma}n_{m\sigma} n_{m^\prime\sigma} + \\
&+& J \sum_{m\neq m^\prime} (d^\dag_{m\up} d^\dag_{m\dn} d_{m^\prime\dn} d_{m^\prime\up} - d^\dag_{m\up} d_{m\dn} d^\dag_{m^\prime\dn} d_{m^\prime\up}).\nonumber
\end{eqnarray}
Here, $n_{m\sigma} = d^\dag_{m\sigma}d_{m\sigma}$, $U$ is the intraorbital interaction and $J$ is the Hund's rule coupling.
The values of $U$ and $J$ have been estimated in Ref. \cite{mravlje2011coherence} using the constrained random phase approximation ($U$=2.3eV and $J$=0.4eV, with $J/U$=0.17)  and in Ref. \cite{PhysRevB.75.035122} using the constrained local density approximation ($U$=3.1eV and $J$=0.7eV, $J/U$=0.23). If the SOC interaction is not included in the calculations, the local multiplet structure is formed by a doublet $\ket{xy\sigma}$ and a quartet  $\{\ket{xz\sigma},\ket{yz\sigma}\}$,  the former having a lower energy. 
Upon inclusion of the SOC in the DFT calculation, the degeneracy of the quartet is broken leading to three doublets,
$\ket{0}, \ket{1}$ and $\ket{2}$. These were calculated by diagonalization of Eq. \ref{eq:obe}, and can be approximately associated with the states $\ket{J=3/2,m_J=\pm1/2}$, $\ket{J=1/2,m_J=\pm1/2}$, and $\ket{J=3/2,m_J=\pm3/2}$, respectively.
The results for the one body energies and the quasiparticle weights are presented in Table \ref{e0_table} for calculations using $U=3.1eV$ and $J/U=0.2$ and $T=5K$~\footnote{We selected this set of parameters, with the largest values for the interactions, for all RISB calculations. This selection partially compensates the RISB failure to fully account for the mass renormalizations. The qualitative features of the solutions are however the same for both sets of parameters and our main conclusions do not depend on this choice.}.
\begin{table}[]
\centering
\begin{tabular}{|c|c|c|c|c|l}
\cline{1-5}
&orbital &{$\mathbf{\varepsilon^0}$ $[\text{eV}]$} & $Z_\mathcal{C}$ & n & \\ \cline{1-5}
\multirow{2}{*}{Without SOC} & $xy$                & -0.44            & 0.61 &1.34 & \\ \cline{2-5}
                            & $xz$, $yz$            & -0.36            & 0.56 &1.33 & \\ \cline{1-5}
\multirow{3}{*}{With SOC}    & 0                 & -0.48            & 0.62 & 1.38 & \\ \cline{2-5}
                            & 1                 & -0.24            & 0.53 & 1.25 &\\ \cline{2-5}
                           & 2                 & -0.42            & 0.58 &1.37 & \\ \cline{1-5}
\end{tabular}
\caption{Diagonal one-body energies ($\mathbf{\varepsilon^0}$), quasiparticle weights ($Z_\mathcal{C}$), and occupancies ($n$) corresponding to Wannier orbitals constructed using the energy window $\mathcal{W}_s=[-3\,\text{eV},1.3\,\text{eV}]$, both with and without including the SOC. Occupancies and quasiparticle weights correspond to the LDA+RISB calculation with parameters $U=3.1\,\text{eV}$ and $J/U=0.2$. }
\label{e0_table}
\end{table}
Although the DFT-DMFT(RISB) results capture the Hund's metal behavior of the system, showing a strong suppression of the charge fluctuations to states with spin other than the maximum, it underestimates the enhancement of quasiparticle renormalization compared to the one obtained by other numerically exact impurity solvers as CTQMC \cite{PhysRevLett.107.256401}. This has been established before \cite{behrmann2012multiorbital} and is a feature shared with other related techniques like the Gutzwiller approximation \cite{PhysRevB.79.075114,PhysRevB.87.045122,PhysRevX.5.011008}, or the slave-spins formalism \cite{PhysRevB.92.075136}. 
One finds moderate enhancement of the quasiparticle mass $\sim 2$ with a slightly smaller enhancement in the $xy$ orbital. This is different to what is obtained using CTQMC where larger mass enhancements $\sim 4$ are found, even at smaller values of the interaction parameters. Additionally, the fact that the mass enhancement in the xy orbital is smaller may suggest that the slave-bosons are more sensitive to the overall bandwidth than to the low-energy fine-structure in the DOS. Later in the text we will see that in spite of this discrepancies the material trends, as the epitaxial strain is changed, are consistent with what is found using CTQMC.

As it was mentioned in Section \ref{sec:intro}, the renormalized masses (with respect to LDA) of the $\al$ and $\be$ sheets of the Fermi surface have been reported to be $\sim 3$ and $\sim 3.5$, respectively \cite{mackenzie2003superconductivity}. 
Note that this experimental finding suggests that, altough possible smaller, there must be additional sources of momentum differentiation of the mass enhacement.
Indeed, the inequality $\frac{m^*}{m_{LDA}}\large|_\al<\frac{m^*}{m_{LDA}}\large|_\be$, or rather $Z_\al>Z_\be$,
is not evident since the Bloch states that conform these sheets have mainly $xz$ and $yz$ symmetry and, as a result,
their weight in the correlated subspace lies on degenerate cubic Wannier orbitals. 
In the following, we show that the spin-orbit coupling enhances the $\vec{k}$ dependence of $\ul{Z}^{QP}_\mathcal{K}(\vec{k})$.

When the SOC is turned on, the doublet $\ket{2}$ is lower in energy than the degenerate orbitals $xz$ and $yz$ in the absence of SOC, as can be observed in Table \ref{e0_table}. Consequently, while $\ket{2}$ increases its occupancy, which drives the orbital away from half-filling and, in turn, gives place to a larger quasiparticle weight, the opposite happens for $\ket{1}$. 
To better compare with the calculation done without SOC, it is naturally convenient to analyze the changes in the basis of cubic Wannier functions.
We find that both the orbital polarization $n_{xy}-n_{xz}$
and the quasiparticle weights $Z^{\mathcal{C}}_{xy}$ and $Z^{\mathcal{C}}_{xz}$ remain essentially constant upon inclusion of the SOC
\footnote{In the cubic harmonic basis we find that the SOC does not change the diagonal elements of the quasiparticle weight, while it originates rather small off-diagonal elements: $Z_{xy\sigma,xz\bar{\sigma}}=0.021$ and $Z_{xz\sigma,yz\sigma}=0.025$. The same occurs with the density matrix, with off-diagonal elements
$n_{xy\sigma,xz\bar{\sigma}}=0.0226$ and $n_{xz\sigma,yz\sigma}=0.0222$.}.
This agrees with the CTQMC results of Ref. \cite{kim2017spin} and supports the conclusion that the SOC does not affect the coherence scale of Sr$_2$RuO$_4$.

\begin{figure}[t]\center
\includegraphics[width=0.5\textwidth,angle=0,keepaspectratio=true]{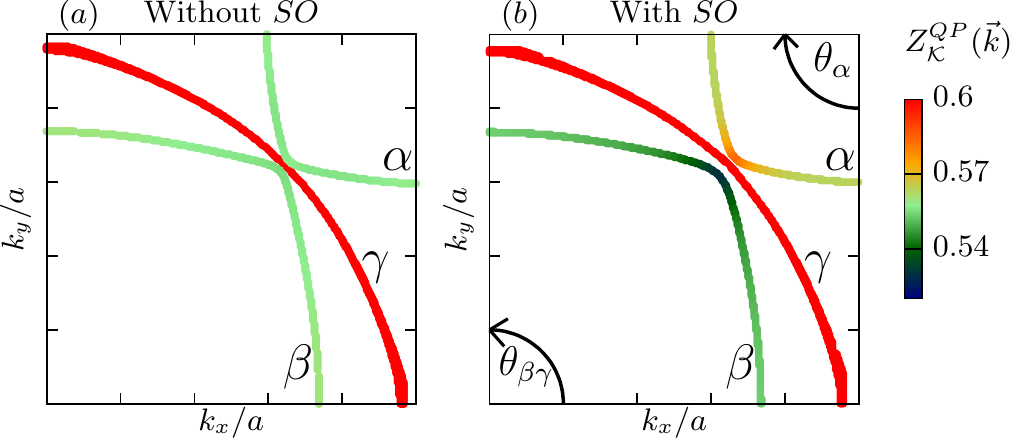}
\includegraphics[width=0.48\textwidth,angle=0,keepaspectratio=true]{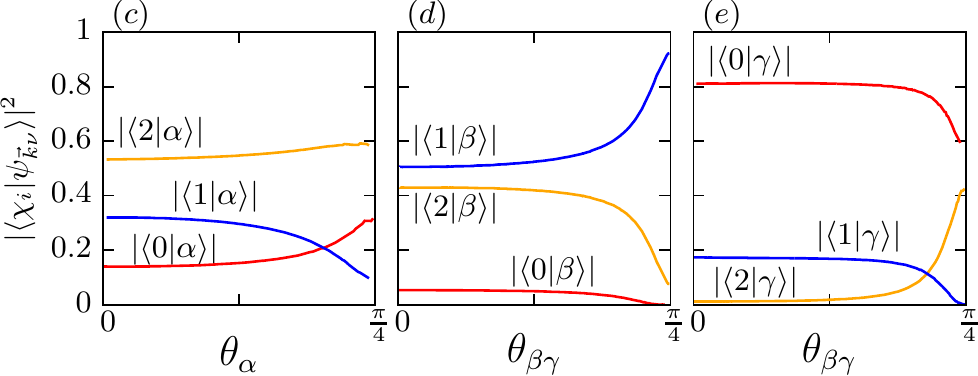}
\caption{Quasiparticle weight on the Fermi surface calculated without including the spin-orbit interaction $(a)$ or including it $(b)$.
$(c)$ Projection of the Bloch states of the $\alpha$ sheet of the Fermi surface on atomic multiplets, as a function of the angle $\theta_\alpha$
defined in $(b)$. $(d)$ and $(e)$ present as a function of the angle $\theta_{\be\gamma}$ the projection on atomic multiplets of the Bloch states from the $\beta$ and $\gamma$ sheets, respectively.
}
\label{zk_sro}
\end{figure}

The SOC does, however, modify the way the quasiparticle weight varies in $\mathcal{K}$ space. 
Figures \ref{zk_sro}(a) and \ref{zk_sro}(b) present the projected quasiparticle weight ${Z}_{\nu_F}^{QP}(\vec{k}_F)$ (for $\nu_F=\alpha,\, \beta,\,\gamma$) [see Eq. (\ref{eq:ZQPFS})] associated with the Fermi surface Bloch states calculated without or with the SOC turned on in the DFT calculations, respectively. 
In the absence of SOC the only dependence on $\vec{k}$ comes from the difference between $Z^{\mathcal{C}}_{xy}$ and $Z^{\mathcal{C}}_{xz}$. 
 Along each Fermi surface sheet the quasiparticle weight is essentially constant.
The inclusion of the SOC leads to a richer momentum differentiation, which is larger at $k_x = k_y$, consistently with the stronger spin-orbital entanglement experimentally observed by spin-resolved ARPES at that line cut of the Brillouin zone \cite{PhysRevLett.101.026406,veenstra2014spin}. 
In particular, the relation $Z_\al > Z_\be$ can only be accounted for in our calculations if the SOC is included.  

The effect of the SOC on $Z^{QP}_{\nu_F}(\vec{k}_F)$ can be understood by analyzing the change in the local multiplets and the subsequent induced charge redistribution. Figs. \ref{zk_sro}(c-e) present the projections of the Bloch states of each sheet onto the different SO multiplets $\ket{0}$, $\ket{1}$ and $\ket{2}$ along the corresponding parametrized angles $\theta_{\al}$ and $\theta_{\be\ga}$, indicated in Fig. \ref{zk_sro}(b).
The $\ga$ band has most of its amplitude on the state $\ket{0}$ along almost the whole range of $\theta_{\be\ga}$.
 The $\al$ sheet has a larger amplitude on the doublet $\ket{2}$, while the $\be$ one has it on $\ket{1}$, specially at $\theta_\alpha=\pi/4$. 
The charge redistribution produced by the SOC, which acts to decrease the quasiparticle weight of $\ket{1}$ while incresing that of $\ket{2}$ contributes, therefore, to obtaining $Z_\al>Z_\be$.

\section{Effects of anisotropic strains}
\label{sec:anisotropic}
Motivated by the recent experimental results reported in Ref. \cite{steppke2016strong} and \cite{burganov2016strain},
in this section we analyze how the electronic correlations evolve in the presence of anisotropic strains.
An interesting question is to what extent the Lifshitz transition, in which the Van Hove singularity (VHS) crosses the
Fermi level, affects the electronic correlations.
We study two situations that have been experimentally addressed and can induce this transition: a biaxial tension that leads to an increase of the lattice parameters $a$ and $b$, and also the response of the system to an uniaxial compression along the [100] direction.
In all cases, we use the experimental lattice parameters $a$, $b$ and $c$ \footnote{A given value of uniaxial (biaxial) deformation fixes $a$ ($a$ and $b$), and we choose the remaining lattice parameters as expected according to the experimental elastic constants reported in Ref. \cite{paglione2002elastic}.} and relax within LDA the internal positions of the apical oxygen and of the transition metal atom.

\begin{figure}[tb]\center
\includegraphics[width=0.48\textwidth,angle=0,keepaspectratio=true]{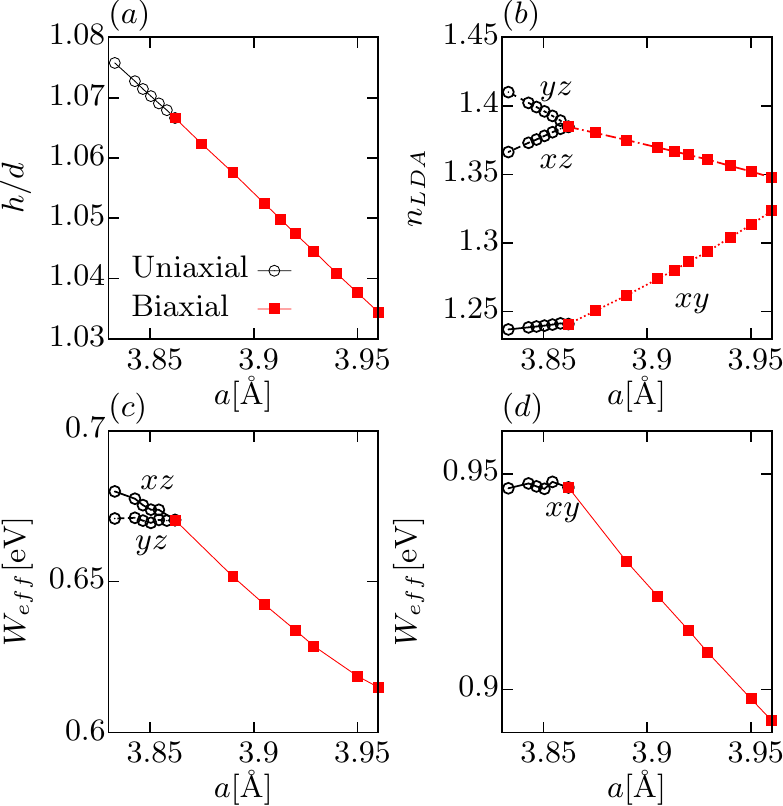}
\caption{a) $h/d$ as a function of $a$, where $h$ is the height of oxygen octahedron and $d=a/2$ the distance betwen the rutenium atom and its first neighbor oxygen. Empty circles and red filled squares correspond to uniaxial and biaxial stress, respectively. b) Occupancy of Wannier orbitals according to LDA. 
	Effective bandwidth of c) $xz$, $yz$  and d)  $xy$ orbitals, as defined in the text. }
\label{occupancy}
\end{figure}

Figure \ref{occupancy}(a) shows the evolution of $h/d$ as a function of $a$, where $h$ is the calculated apical oxygen height and $d$=$a/2$ the distance between the Ru and its next oxygen neighbors. 
For both uniaxial and biaxial strains, $h/d$ evolves towards a more regular octahedron, from 1.07 to 1.03. In the biaxial case, consequently, the splitting between the t$_{2g}$ bands diminishes. That is, the on-site $xy$ energy decreases while the  $\{xz,yz\}$ increases, contributing to transfer charge from the  $\{xz,yz\}$  orbitals to the $xy$ one, as confirmed by the evolution of the corresponding occupancies [see Fig. \ref{occupancy}(b)]~\footnote{It is important to note that contrary to what is expected from a pure electrostatic model for a $h/d$>1 octahedral environment, the $xy$ on-site energy is lower than the $\{xz,yz\}$ ones (see Table \ref{e0_table}). In Ref. \cite{0953-8984-27-17-175503} possible reasons for this negative splitting are discussed.}.
The uniaxial compression mainly transfers charge from $xz$ to $yz$, the reason being the increased level energy of the $xz$ orbital
due to the shorter distances along $x$.

In the presence of a VHS, it is convenient to analyze the strength of the correlations in terms of an effective bandwidth W$_{eff}$ calculated through the second moment of the density of states, as suggested in Ref. \cite{PhysRevB.80.245112}. The calculated  W$_{eff}$ for the $t_{2g}$ bands are shown in Figs. \ref{occupancy}(c) and \ref{occupancy}(d). 
It can be observed that, even for the $xy$ orbital, the W$_{eff}$ decreases smoothly with $a$.
The relative change is only slightly larger in the $\{xz,yz\}$ bands than in the $xy$ one. 

In the following, we present how the electronic correlations, as measured by the quasiparticle weight, evolve under the two kinds of deformation mentioned above.

\subsection{Biaxial strain}

Here we present results for strained Sr$_2$RuO$_4$ obtained with the formalism introduced in Sec. \ref{sec:methods} and DMFT calculations using CTQMC as impurity solver.

\begin{figure}[t]\center
\includegraphics[width=0.48\textwidth,angle=0,keepaspectratio=true]{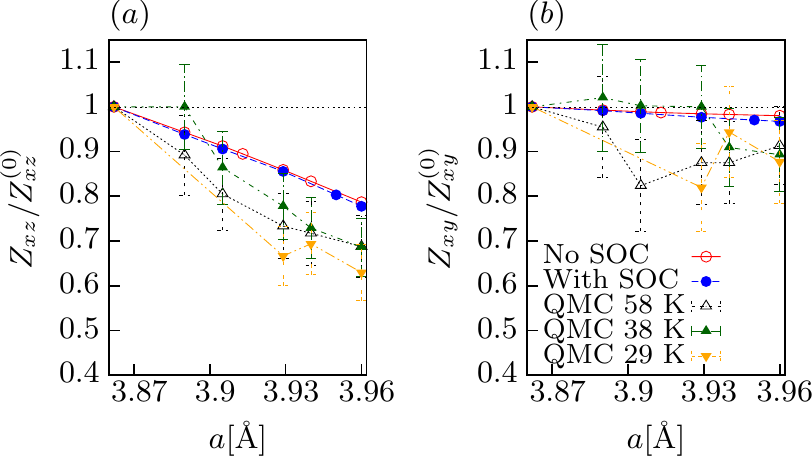}
\includegraphics[width=0.46\textwidth,angle=0,keepaspectratio=true]{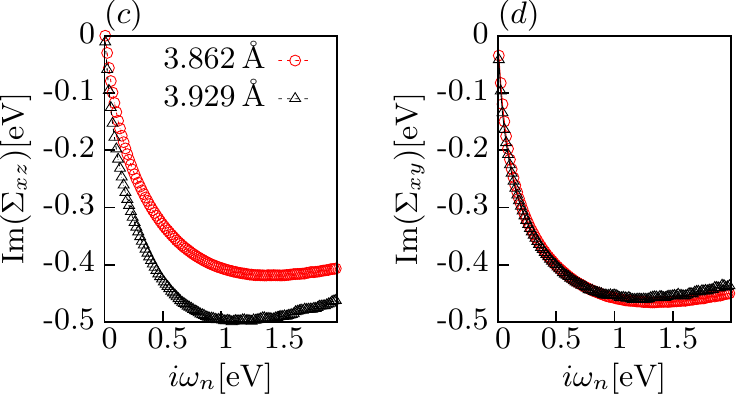}
\caption{Quasiparticle weight of the $\{xz,yz\}$ $(a)$ and $xy$ orbitals $(b)$ as a function of lattice parameter $a$ calculated using RISB at $T=5K$. 
Empty red and filled blue symbol datapoints are obtained without and with SOC, respectively. Triangle symbols are CTQMC results without SOC. 
(c-d) Imaginary part of the self-energy calculated with CTQMC at different temperatures. The interaction parameters are $U=2.3\,\text{eV}$ and $J=0.4\,\text{eV}$.
}
\label{Z_vs_a}
\end{figure}

Figures \ref{Z_vs_a}(a) and \ref{Z_vs_a}(b) show the evolution of the quasiparticle weight associated with the $\{xz,yz\}$ and $xy$ orbitals, respectively, relative to the corresponding values for the unstressed compound (noted as $Z^{(0)}_\al$), as a function of the lattice parameter $a$. The RISB results depend very weakly on the temperature for $T\lesssim 30K$. 
The inclusion of the SOC introduces small changes in the quasiparticle weight. In the range of lattice parameters considered, the observed reduction upon including the SOC is less than 1\%.

Both $Z_{xy}$ and $ Z_{xz,yz}$ decrease monotonically as $a$ increases from the unstrained case, in line with the monotonic reduction of W$_{eff}$ shown in Figs. \ref{occupancy}(c) and \ref{occupancy}(d).
The smaller variation rate of $Z_{xy}$ can be associated with two effects. First, the bandwidth reduction of the xy band is percentually smaller than that of the \{xz,yz\} bands. Second, there is a compensating effect in the correlations generated by the increase in the occupancy of the $xy$ orbital with $a$ which drives it further away from half-filling.

We also performed CTQMC calculations for Sr$_2$RuO$_4$ under biaxial strain. For these calculations we used the parameters $U$=2.3eV and $J$=0.4eV.  The obtained quasiparticle weights, presented in Fig. \ref{Z_vs_a}, indicate, similarly to the RISB results, that the biaxial stress has a stronger effect on the $xz$, $yz$ orbitals. 
The mass enhancement is computed using a 4th order polynominal fit to the lowest six Matsubara points of the self-energy shown in Fig. \ref{Z_vs_a}(c-d).
Typically, the values of mass renormalization fluctuate by the order of $10\%$ from DMFT iteration to iteration, hence we present the data with such errorbar.
At $T=29\,K$, the effective mass of xz--yz orbitals presents an increment relative to the unstressed compound of $\sim35\%$ (from 3.4 to 4.6),
while that of the xy orbital is of the order of the statistical error, $\sim10\%$.
\begin{figure}[t]\center
\includegraphics[width=0.48\textwidth,angle=0,keepaspectratio=true]{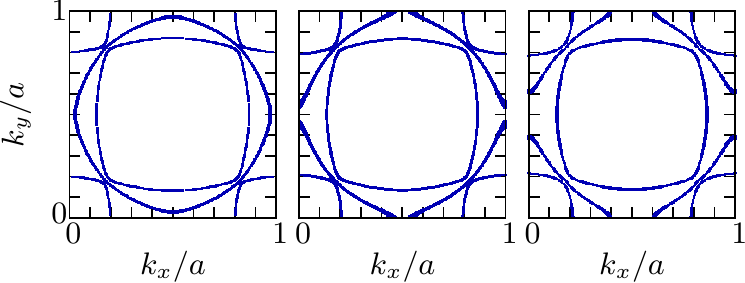}
\caption{Evolution of the renormalized Fermi surfaces (calculated with SOC) for $a$= 3.862, 3.890, 3.929$\,\text{\AA}$ (from left to right) showing that the system undergoes a Lifshitz transition close to $a$= 3.890 $\text{\AA}$.  
}
\label{Lifshitz}
\end{figure}

It is important to point out that the range of values of the lattice parameter $a$ studied, includes the Lifshitz transition of this compound.
Figure \ref{Lifshitz} shows the Fermi surfaces calculated with RISB including the SOC for three different values of $a$= 3.862, 3.890, 3.929$\text{\,\AA}$.  It can be observed that while the electron--like $\beta$ and the hole--like $\alpha$ sheets shrink with increasing $a$, the $\gamma$ one (having mainly $xy$ symmetry) undergoes a Lifshitz transition,  changing its character from electron--like to hole--like. This theoretical result agrees with the experimental data reported in Ref. \cite{burganov2016strain},  
obtained for thin films grown on top of different substrates.

The Lifshitz transition occurs already at the LDA level~\cite{hsu2016manipulating}. The effect of an improved treatment of the correlations through RISB is to slightly decrease the value of $a$ at which the transition takes place ($a_c$). More precisely, LDA gives $a_c$=3.98 $\text{\AA}$, RISB $a_c$=3.91 $\text{\AA}$ and RISB with SOC$a_c$=3.89 $\text{\AA}$. The value of $a_c$ naturally depends on  U because this parameter directly affects the occupancy of the $xy$ orbital.

The main conclusion to point out from these results is that there is no significant effect on the electronic correlations, as measured by $Z$, associated with the occurrence of the Lifshitz transition (see also Ref. \cite{PhysRevB.93.155103}). This observation is in qualitative agreement with the experimental results of Ref.  \cite{burganov2016strain}.

\subsection{Uniaxial strain}

In this section, we study the evolution of the correlation strength as a function of the uniaxial compressive stress $\epsilon_{xx}$. As mentioned before,  we use the experimental lattice parameters $a$, $b$ and $c$ \cite{steppke2016strong} and relax the internal positions.
The results are expected to be symmetric with respect to tensile stress.

\begin{figure}[tb]\center
\includegraphics[width=0.48\textwidth,angle=0,keepaspectratio=true]{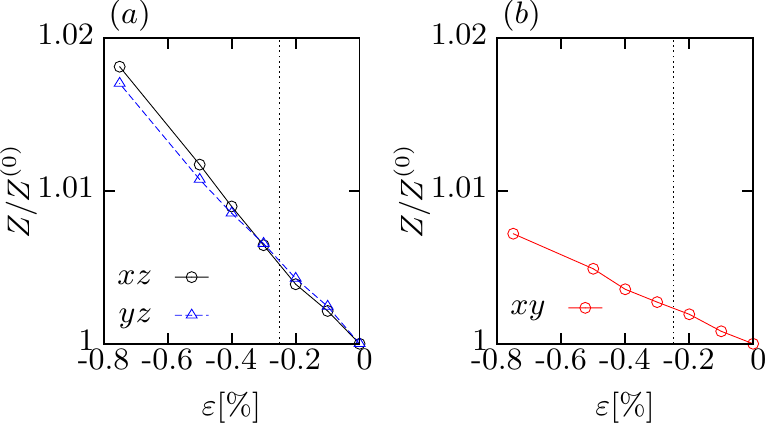}
\includegraphics[width=0.48\textwidth,angle=0,keepaspectratio=true]{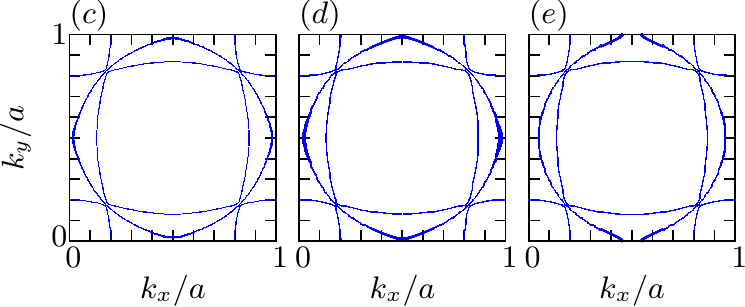}
\caption{$(a)$ Quasiparticle weight of the orbital $\{xz,yz\}$ as function of the stress $\epsilon_{xx}$ at $T=5\,$K.
	$(b)$ Quasiparticle weight of the orbital $xy$ as function of the stress $\epsilon_{xx}$. c-e) Evolution of the renormalized Fermi surfaces for the percentage uniaxial stress $\epsilon_{xx}$= 0, -0.2 y -0.5$\%$ (from left to right) showing that the system undergoes a Lifshitz transition close to $\epsilon_{xx}$=0.2$\%$ around the $(\frac{\pi}{2a}, 0)$ point.
}
\label{data_uniaxial}
\end{figure}

Overall, the effect of uniaxial stress on Z is much weaker than in the biaxial case, basically because the induced [100] relative distortion is smaller (up to 0.8 $\%$).
On average, the quasiparticle weight of the $t_{2g}$ states slightly increases with uniaxial pressure. Moreover, as the case of biaxial distortion, the correlation strength evolves monotonously through the Lifshitz transition. 
Figures \ref{data_uniaxial}(a) y \ref{data_uniaxial}(b) show the evolution of the quasiparticle weigth corresponding to the $xz$, $yz$  and $xy$ states, respectively. We present only the RISB results without SOC, since the effect of SOC is negligible for these small values of compression. 

It can be observed that, again, the variation rate of $Z_{xy}$ is much smaller than that of $ Z_{xz,yz}$. This is consistent with the almost constant behaviour of the corresponding effective 
bandwidth and occupancy under uniaxial strain in the former case [see Fig. \ref{occupancy} (d)].

\section{Conclusions}
\label{sec:conclusion}
In this article we presented an implementation of the multiorbital slave-boson rotationally invariant (RISB) quantum impurity solver within the LDA+DMFT approach to {\it ab initio} calculations of strongly correlated systems. The main disadvantage of the RISB solver is that in general it can only be trusted at a qualitative level and that it fails to describe non-Fermi liquid behaviors besides trivial insulating phases. Its main advantages, compared to other numerically exact solvers as Quantum Monte Carlo, are its much lower computational cost at low temperatures which allows to rapidly explore a wide range of parameters and materials, and the abilty of easily handling off-diagonal hybridizations at arbitrary low temperatures in problems that in general give place to a sign problem in CTQMC. It can therefore be used as a tool to explore and identify correlated regimes worth of a more detailed study using other techniques as CTQMC or numerical renormalization group approaches.
An important result of our implementation is that we obtained transformation relations between physical quantities, as the quasiparticle mass renormalization, in the local correlated space and their counterparts in Bloch space. This transformation allows to obtain the momentum dependence of the quasiparticle mass renormalization independently of the quantum impurity solver used to treat the DMFT equations.
We applied the RISB approach to study the electronic correlations, as measured by the quasiparticle mass enhancement, of Sr$_2$RuO$_4$. We found that it is necessary to include the spin-orbit coupling in the DFT calculations to explain the experimentally observed mass enhancement differentiation between the $\alpha$ and $\beta$ sheets of the Fermi surface. We also find that the SOC strongly enhances the momentum dependence of the quasiparticle weight on both sheets while the $\gamma$  sheet is largely unaffected.

A biaxial stretching of the compound on the $a-b$ plane leads to a monotonic increase of the mass enhancement that is larger for the $xz$ and $yz$ orbitals, which determine the mass enhancement of the $\alpha$ and $\beta$ Fermi sheets, than for the $xy$ orbital, which determines the mass enhancement of the $\gamma$ Fermi sheet. While the sample suffers a Lifshitz transition for $a=b\sim 3.9$\AA, 
we did not find any significant change of behavior of the mass enhancement across it.
We also performed calculations using the numerically exact CTQMC solver which confirmed the absence of any dramatic effect of the electronic correlations at the Lifshitz transition. 

Finally, we analyzed the behavior of the mass enhancement when the sample is under uniaxial strain. As in the biaxial case, we found a monotonic behavior with no significant features across the Lifshitz transition.         

\acknowledgments
This work was partly supported by program ECOS-MINCyT France-Argentina under project A13E04 (J.F, L.P., V.V, P.C).
J.F and P.C. acknowledge support from SeCTyP UNCuyo Grant 06/C489 and PICT- 201-0204 ANPCyT.
V.V. thanks also financial support from ANPCyT (PICT 20150869, PICTE 2014 134), CONICET (PIP 00273/12 GI).
This work was also supported by the European Research Council grants ERC-319286-QMAC (J.M., L.P.). J.M. acknowledges support by the Slovenian Research Agency (ARRS) under Program P1-0044.

\bibliography{references}

\appendix

\section{Inverse relations} \label{app:inverse}

The inverse of Eq. (\ref{Znu})  can be obtained by applying $\ul{P}(\vec{k})$ and $\ul{P}^\dag(\vec{k})$ at left and right, respectively, and using $\ul{P}(\vec{k}) \ul{P}^\dag(\vec{k}) = \ul{1}_{\mathcal{C}}$, it reads:
\begin{equation}
\label{eq:zimp}
\ul{P}(\vec{k}) \ul{Z}^{-1}_\mathcal{K}(\vec{k}) \ul{P}^\dag(\vec{k}) = \ul{Z}_\mathcal{C}^{-1}
\end{equation}
The asymmetry between Eqs. (\ref{Znu}) and (\ref{eq:zimp}) comes from the fact that the transformations $\ul{P}(\vec{k})$ are not unitary in the general case. 

Based on Eq. (\ref{Znu}), we define:
\begin{equation}
\label{eq:rk}
\ul{R}^{-1}_\mathcal{K}(\vec{k}) = \idkk-\ul{P}^\dag(\vec{k})\ul{P}(\vec{k}) + \ul{P}^\dag(\vec{k}) \ul{R}_\mathcal{C}^{-1}\ul{P}(\vec{k})
\end{equation}
and in analogy with Eq. (\ref{eq:zimp}) we obtain:
\begin{align}
\label{eq:rimp1}
\ul{P}(\vec{k}) \ul{R}^{-1}_\mathcal{K}(\vec{k}) \ul{P}^\dag(\vec{k}) &= \ul{R}_\mathcal{C}^{-1}. 
\end{align}
From Eq. (\ref{eq:rk}) it can be shown by direct calculation that the quasiparticle mass renormalization in $\mathcal{K}$ space satisfies 
\begin{equation}
    \ul{Z}_\mathcal{K}(\vec{k})=\ul{R}_\mathcal{K}(\vec{k}) \ul{R}_\mathcal{K}^\dag(\vec{k}).
    \label{eq:ZKRR}
\end{equation}


The relation between transformation matrices given by Eq. (\ref{eq:rk}) is consistent
with the relations between quasiparticle and physical Green's functions in both $\mathcal{K}$ and $\mathcal{C}$ spaces. 
To show this we transform Eq. (\ref{eq:relation}) to $\mathcal{C}$ space.
We first derive the following relation between the transformation matrices in $\mathcal{C}$ and $\mathcal{K}$.
\begin{equation}
\label{eq:inv1}
\ul{P}(\vec{k}) \ul{R}_\mathcal{K}^{-1}(\vec{k}) = \ul{R}_\mathcal{C}^{-1} \ul{P}(\vec{k}),
\end{equation}
by applying $\ul{P}(\vec{k})$ to the left of Eq. (\ref{eq:rk}). Similarly, we also obtain:
\begin{equation}
\label{eq:inv2}
\ul{R}_\mathcal{K}^{\dag-1}(\vec{k})\ul{P}^\dag(\vec{k}) = \ul{P}^\dag(\vec{k}) \ul{R}^{\dag-1}_\mathcal{C}.
\end{equation}
From Eq. (\ref{eq:relation}) it follows that:
\begin{equation}
\label{eq:Gqp}
\ul{G_{qp}}(\vec{k},\iom) = \ul{R}_\mathcal{K}^{-1}(\vec{k}) \ul{G}(\vec{k},\iom) [\ul{R}_\mathcal{K}^\dag]^{-1}(\vec{k}).
\end{equation}
We project this equation to $\mathcal{C}$ multiplying at left by $\ul{P}(\vec{k})$, at right by $\ul{P}^\dag(\vec{k})$ and summing over $\vec{k}$:
\begin{align}
&\ul{G_{qp}^{loc}}(\iom) = \sum_{\vec{k}} \ul{P}(\vec{k}) \ul{R}_\mathcal{K}^{-1}(\vec{k}) \ul{G}(\vec{k},\iom)\ul{R}_\mathcal{K}^{\dag-1}(\vec{k}) \ul{P}^\dag(\vec{k}).  \nonumber
\end{align}
Using Eqs. (\ref{eq:inv1}) and Eq. (\ref{eq:inv2}), we have:
\begin{align}
 &\ul{G_{qp}^{loc}}(\iom) = \sum_{\vec{k}} \ul{R}_\mathcal{C}^{-1} \ul{P}(\vec{k}) \ul{G}(\vec{k},\iom) \ul{P}^\dag(\vec{k}) \ul{R}^{\dag-1}_\mathcal{C} \nonumber \\
                 &= \ul{R}_\mathcal{C}^{-1}\Big( \sum_{\vec{k}} \ul{P}(\vec{k}) \ul{G}(\vec{k},\iom) \ul{P}^\dag(\vec{k})\Big)\ul{R}^{\dag-1}_\mathcal{C}  \nonumber \\
                 &= \ul{R}_\mathcal{C}^{-1} \ul{G}^{loc}(\iom) \ul{R}^{\dag-1}_\mathcal{C},
\end{align}
and finally
\begin{align}
\ul{R}_\mathcal{C} \ul{G_{qp}^{loc}}(\iom)  \ul{R}^\dag_\mathcal{C} = \ul{G^{loc}}(\iom),
\end{align}
which is the expected relation between the physical and the quasiparticle Green's functions in $\mathcal{C}$.

\section{Derivation of Eq. (\ref{eq:gqp}).}
\label{ap_gqp}
The quasiparticle lattice Green's function can be computed using the DFT+DMFT equations and the self-energy obtained in the RISB saddle-point approximation.
To this aim, we substitute in Eq. (\ref{eq:gkw}) the self-energy given by Eq. (\ref{eq:sigmaRISB}). 
The terms linear in frequency and in the chemical potential read (in the following we omit the dependence in $\vec{k}$ of $\ul{P}$ and $\ul{R}_\mathcal{K}$):
\begin{equation}
(\iom + \mu) [\idkk-\ul{P}^\dag(\idc - \ul{Z}_\mathcal{C}^{-1})\ul{P}] = (\iom + \mu) \ul{Z}^{-1}_\mathcal{K},
\end{equation}
where we have used Eq. (\ref{Znu}). The term proportional to $\mathbf{\Lambda}_\mathcal{C}$ reads:
\begin{equation}
\ul{P}^\dag\ul{R}^{\dag-1}_\mathcal{C} \ul{\Lambda}_\mathcal{C}  \ul{R}^{-1}_\mathcal{C}\ul{P} = \ul{R}^{\dag-1}_\mathcal{K}\ul{P}^\dag \ul{\Lambda}_\mathcal{C} \ul{P} \ul{R}^{-1}_\mathcal{K},
\end{equation}
where the equality follows using Eqs. (\ref{eq:inv1}) and (\ref{eq:inv2}).
The other terms are:
\begin{equation}
-\ul{\varepsilon}(\vec{k})+\ul{P}^\dag\big(\ul{\varepsilon^0}+\ul{\Sigma^{dc}}\big)\ul{P}.
\end{equation}
The summation of these terms and use of Eq. (\ref{eq:relation}) leads to Eq. (\ref{eq:gqp}).

\section{Application to SrVO$_3$}
\label{svo_results}

\begin{figure}[ht]
\center
\includegraphics[width=0.4\textwidth,angle=0,keepaspectratio=true]{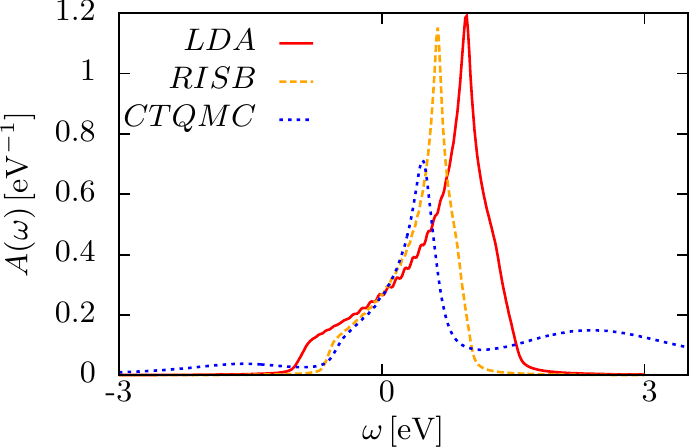}
\caption{Density of states projected on the t$_2g$ Wannier orbital, calculated using LDA, LDA+DMFT(RISB), and LDA+DMFT(CTQMC)}
\label{svo_ws}
\end{figure}

Here we present as a benchmark a comparison between results obtained with LDA+RISB and LDA+DMFT (CTQMC) for the compound SrVO$_3$.
We use values of $U=4.5\,$eV and $J=0.6\,$eV, and construct Wannier functions for the $t_{2g}$ bands within 
the energy window $[-1, 2]\,$eV.
Fig. \ref{svo_ws} presents the spectral density as function of the energy for each of the three-fold degenerated $t_{2g}$ Wannier orbitals. 
It can be observed that the RISB method provides a reasonably good approximation at low energies in spite of subestimating 
the bandwith renormalization. The values of $Z$ obtained are $\sim0.5$ and $\sim 0.65$ for CTQMC and RISB, respectively.

\end{document}